\documentstyle[12pt]{article}
\newcommand{\ben}{\begin{enumerate}}
\newcommand{\een}{\end{enumerate}}
\newcommand{\be}{\begin{equation}}
\newcommand{\ee}{\end{equation}}
\newcommand{\bse}{\begin{subequation}}
\newcommand{\ese}{\end{subequation}}
\newcommand{\bea}{\begin{eqnarray}}
\newcommand{\eea}{\end{eqnarray}}
\newcommand{\bc}{\begin{center}}
\newcommand{\ec}{\end{center}}

\newcommand{\CR}{\nonumber \\}

\newcommand{\A}{\alpha}

\newcommand{\lm}{\lambda}

\newcommand{\E}{\epsilon}

\newcommand{\SL}{\ ^{*} {\cal L}}

\newcommand{\RE}{{\cal R}}
\newcommand{\SR}{\ ^{*} {\cal R}}
\newcommand{\NA}{{\cal N}}
\newcommand{\SN}{\ ^{*} {\cal N}}
\newcommand{\SM}{\ ^{*} {\cal M}}
\newcommand{\MON}{{\rm Mon}(r|\SL)}
\newcommand{\MONP}{{\rm Mon}(r'|\SL)}
\newcommand{\SE}{\ ^{*}\varepsilon}

\begin{document}

\begin{center}

{\large {\bf Field Theory on Infinitesimal-Lattice Spaces }}

\vskip30pt

      Tsunehiro KOBAYASHI 
      \vskip5pt
      
      {\it Department of General Education, 
       Tsukuba College of Technology, 
       
             Tsukuba, 305-0005 Ibaraki, Japan }
      
      E-mail: kobayash@a.tsukuba-tech.ac.jp
\end{center}

\begin{abstract}
Equivalence in physics is discussed on the basis of experimental data 
accompanied by experimental errors. 
It is pointed out that the introduction of the equivalence being consistent 
with  the mathematical definition 
is possible only in theories constructed 
on non-standard number spaces 
by taking the experimental errors 
as infinitesimal numbers. 
Following the idea for the equivalence, 
a new description of space-time $\SL$ in terms of infinitesimal-lattice points 
on non-standard real number space $\SR$ is proposed. 
By using infinitesimal neighborhoos ($\MON$) of real number $r$ on $\SL$ 
we can make a space $\SM$ which is isomorphic to $\RE$ as additive group. 
Therefore, every point on $(\SM)^N$ automatically has 
the internal confined-subspace $\MON$. 
A field theory on $\SL$ is proposed. 
It is shown that $U(1)$ and $SU(N)$ symmetries 
on the space $(\SM)^N$ are induced from the internal substructure 
$(\MON)^N$. 
Quantized state describing configuration space is constructed on $(\SM)^N$. 
We see that 
Lorentz and general relativistic transformations are also represented by 
operators which involve the $U(1)$ and $SU(N)$ internal symmetries. 
\end{abstract}

\section{Introduction}
We would like to start from a question, 
\bc 
{\it ``Why are non-standard spaces needed in theories of physics?''}
\ec
In our observations the judgment of equivalence between two or more phenomena 
plays a very important role. 
It is kown that the equivalence is rigorously defined 
in mathematics in terms of the following 
three conditions; 

(1) $A \sim A\, $ (reflection)

(2) $A \sim B \Longrightarrow B \sim A,$ (symmetry)

(3) $A \sim B,\ \ B \sim C \Longrightarrow A \sim C. $ (transitivity)
\hfil\break
In observations of physics, that is, in experiments, 
the equivalence (physical equivalence) can be described as follows:
\begin{center}
{\it Two phenomena {\rm A} and {\rm B} are equivalent, 

if {\rm A} and {\rm B} coincide within the experimental errors.}
\end{center}
It should be stressed that the physical equivalence is detemined by the experimental errors. 
Futhermore we must recognize that there is no experiments accompanied by 
no error. 
We should consider that experimental errors are one of the fundamental 
observables in our experiments. 
It is quite hard to understand that there is no theory which involves 
any description of experimental errors, even though they 
are very fundamental observables. 
It is also hard to understand that 
the question whether such physical equivalence is compatible 
with the mathematical definition represented by the above three 
conditions had never been discussed.  
Let us discuss the question here. 
We easily see that the first two conditions, that is, reflection and symmetry 
are compatible with the physical equivalence based on experimental errors. 
We can, however, easily 
present examples which break the third condition (transitivity), 
that is to say, 
$A\sim B$ and $B\sim C$ are satisfied within their errors 
but $A$ and $C$ does not coincide within their errors. 
This arises from the fact that real numbers which exceed any real numbers 
can be made from repeated additions of a non-zero real number 
because of Archimedian property of real number space. 
\bc
``{\it How can we introduce the physical equivalence in theories?''} 
\ec 
Consistent definition of the physical equivalence is allowed, only when 
experimental errors are taken as {\it infinitesimal numbers}[1] 
in non-standard spaces. 
This result comes from the fact that any non-zero real numbers cannot be made 
from any finite sum of infinitesimal numbers. 
Any repetitions of the transitivity, that is, repeated additions of 
any infinitesimal numbers does not lead any non-zero 
real numbers. 
We can describe the situation as follows;
$$
\forall\E \in {\rm Mon}(0)\ \ {\rm and}\ \ \forall N \in {\cal N} 
\Longrightarrow \E N \in {\rm Mon}(0),
$$
where Mon(0) and ${\cal N}$, respectively, stand for the set of all 
infinitesimal numbers on non-standard spaces and the set of all natural 
numbers. 
From the above argument we can conclude that 
we must make theories, in which the physical equivalence based on 
experimental errors is described in terms of 
the mathematically consistent form, on a non-standard space. 
This is the reason why non-standard spaces are needed in the description of 
realistic theories based on the physical equivalence. 
It is once more stressed that such realistic theories 
must involve the fundamental observables, experimental 
errors, in the mathematically rigorous way. 

An example for the introduction of the physical equivalence in quantum 
mechanics on non-standard space has been presented in the derivation 
of decoherence between quantum states for the description of quantum theory 
of measurements.[2-4] 
Though we have many other interesting problems for the construction of 
theories 
on non-standard spaces,[5-14] we shall investigate space-time structure and 
field theory in this paper.

Following the above argument, let us dicuss about 
observation of continuity of space-time. 
Whether space-time is continuous (as represented by the set of 
real numbers $\RE$) or discrete (as represented by the set of discrete 
lattice-points) 
is a fundamental question for the space-time structure. 
We may ask 
\bc
``{\it How can we experimentally verify the continuous property of space-time?''. }
\ec 
Taking into account that 
experimental errors are fundamental observables in physical phenomena, 
we should understand that the continuity of space-time cannot be directly 
verified in any experiments. 
This means that a discrete space-time is sufficient to describe realistic 
space-time. 
We, however, know that translational and rotational invariances (including 
Lorenz invariance) with respect to
space-time axes seems to be very fundamental 
concepts in nature and lattice spaces break them. 
This disadvantage seem to be very difficult to 
overcome on usual lattice spaces having a finite 
lattice-spacing between two neighboring lattice-points. 
As noted in the above argument, 
experimental errors must be described in terms of infinitesimal numbers on 
non-standard spaces. 
On non-standard spaces[1] we can introduce infinitesimal lengths 
which are smaller than all real numbers except $0$. 
It will be an interesting question whether we can overcome the disadvantage on 
lattice spaces defined by infinitesimal lattice-spacing. 
Actually such infinitesimal discreteness cannot be observed in our experiments, 
where all results must be described by real numbers. 
This fact indicates that such lattice space-time will possibly be observed 
as continuous structure. 
Hereafter we call lattice spaces discretized by infinitesimal numbers 
{\it infinitesimal-lattice spaces} and they are denoted by $\SL$.[11] 
That is to say, 
such a lattice space $\SL$ is constructed as the set of non-standard numbers 
which are separated by an infinitesimal lattice-spacing $\SE$
on $\SR$(the non-standard extension  of $\RE$). 
It is transparent that such $\SL$ do not contain many of real numbers 
in general.  
There is, however, a possibility that parts of infinitesimal neighborhoods of 
all real numbers are contained in $\SL$, because it is known that 
the power of $\SL$ is  same as that of $\RE$.[1] 
Thus there is a posibility 
that a space constructed from the set of all 
infinitesimal neighborhoods on $\SL$ will be isomorphic to $\RE$ 
and translations and rotations on the space can be introduced as same as 
those on $\RE$.[11] 
We shall start from the investigation of 
properties of $\SL$ and examine the construction of 
a new theory on the space-time represented by $\SL$, 
where the space-time are not treated as parameters but written by operators.
[12,13,14]

\section{ Infinitesimal-lattice spaces $\SL$}
Let us take a non-standard natural number 
$
^*N\in\ \SN-\NA,
$ 
which is an infinity.[1] 
We take the closed set $[-\ ^*N/2,\ ^*N/2]$ on $^*\RE$ and put $(\ ^*N)^2
\ +\ 1$
points with an equal spacing $\SE=\ ^*N^{-1}$ on the set. 
For the convenience of the following discussions $\ ^*N$ is chosen as 
$\ ^*N/2 \in\ ^*{\cal N}$. 
The length between two neighboring points is $\SE$ which is 
an infinitesimal, i.e. $\ ^*\E \in$Mon(0). 
Let us consider the set of the infinitesimal lattice-points $\SL$,[11] 
which consists 
of these $(\ ^*N)^2\ +\ 1$ discrete points on the closed set. 
Lattice-points on $\SL$ are written by 
$
l_n=n  \SE, 
$ 
where $n\in \ ^*{\cal Z}$ 
and fulfil the relation 
$
-(\ ^*N)^2/2\leq n \leq (\ ^*N)^2/2.
$ 
From the process of the construction of $\SL$ it is transparent that 
$
\SL\not\supset \RE.
$ 
Actually it is obvious that all irrational numbers of $\RE$ are not contained 
in $\SL$, because $\ ^*N$ is taken as an element of $\SN$ and 
$n\ ^*\E=n/\ ^*N$ is an element of $\ ^*{\cal Z}$. 

Let us show a theorem:
\bc
{\it Monads of all real numbers, {\rm Mon}($r$) $\forall r \in \RE$, 
have their elements on $\SL$. }
\ec 
Proof: Take a real number $r\in \RE$. 
The number $r$ is contained in the closed set $[-\ ^*N/2,\ 
^*N/2]$ on $\SR$, because $^*N$ is an infinity of $\SN$ and then 
$[-\ ^*N/2,\ ^*N/2]\supset \RE$. 
Since the lattice-points of $\SL$ divide the closed set
into $( ^*N)^2$ regions of which 
lenght is $\SE$, the real number $r$ must be on a lattice-point or between 
two neighboring lattice-points whose distance is $\SE$. 
We can, therefore, find out a non-standard integer $N_r$ fulfilling the 
following relation;
\be
N_r \SE\leq r< (N_r+1) \SE,
\ee 
where $|N_r|\in\SN-\NA$. 
The difference $r-N_r \SE$ is an infinitesimal number smaller than $\SE$. 
Thus we can define the infinitesimal neighborhood of $r$ on $\SL$ such that 
\be
{\rm Mon}(r|\SL)\equiv \{ l_n(r)=(N_r+n)\SE|n\in \ ^*{\cal Z},\ n\SE\in {\rm Mon}(0)\}. 
\ee
The relation of the standard part map[1] 
$
{\rm st}(l_n(r))=r
$ 
is obvious. 
The theorem has been proved. 
Hereafter we shall call Mon($r|\SL$) and its elements $l_n(r)$ 
monad lattice-space ($\SL$-monad) and monad lattice-points, respectively. 

From the above argument we see that there is one-to-one correspondence between 
$\RE$ and 
$ 
\SL_{l(\RE)}\equiv \{ l_0(r)|r\in \RE\}
$ 
(the set of $l_0(r)$ 
for $\forall r\in \RE$) with respect to the correspondence $r\leftrightarrow
l_0(r)$. 
Note also that from the definition of monad we have the relations 
\be 
\MON \cap \MONP=\phi,\ \ \ {\rm for}\ \ r\not=r',\ r,r'\in \RE.
\ee 

Magnitudes of lattice-points contained in all of the monad lattice-space 
Mon($r|\SL$) for $\forall r\in \RE$ are not infinity,  
because they are elements of monads of real numbers. 
We shall write the set of all these finite lattice-pionts by 
$$
\SL_\RE\equiv \{ l_n(r)|r\in\RE,\ n\in\ ^*{\cal Z},\ n\SE\in {\rm Mon}(0)\}= 
\cup_{r\in \RE}\ \MON. 
$$
The sets $\SL_\RE$ and Mon$(0|\SL)$ are additive groups. 
Note here that $\SL_{l(\RE)}$ is not an additive group, 
because in general $l_0(r)+l_0(r')\not=l_0(r+r')$ possibly happens, 
that is, $N_{r+r'}$ is not always equal to $N_r+N_{r'}$
but possibly equal to $N_r+N_{r'}+1$.
It is apparent that 
$
  \SL_\RE=\SL_{l(\RE)}+{\rm Mon}(0|\SL)\ \ {\rm and}\ \  
\SL_{l(\RE)}\cap {\rm Mon}(0|\SL)=\{ 0\}. 
$  
Let us introduce the quotient set of $\SL_\RE$ by ${\rm Mon}(0|\SL)$
as 
\be
\SM\equiv \SL_\RE/{\rm Mon}(0|\SL).
\ee 
From one-to-one correspondence between $\RE$ and $\SL_{l(\RE)}$ 
we see that there is one-to-one correspondence 
between $\RE$ and $\SM$, and thus 
$
\SM \cong \RE
$ 
as additive groups, where the addition on $\SM$ may be described
by st-map of the addition  on $\SL_\RE$ such that 
$ {\rm st}(l_n(r)+l_m(r'))=r+r'$ 
for $\forall l_n(r)\in\MON$ and $\forall l_m(r')\in \MONP$ with $r,r'\in \RE$. 

We can introduce translations and rotations on $\SM$ by using the relation 
$
\SM \cong \RE.
$ 
 (See refs. 11 and 14.)

\section{ Confined fractal-like property of $\SL$}
Though we have shown that $\SM\cong \RE$, 
there is a large difference between them, that is, 
$\SM$ is constructed from the monad lattice-spaces $\MON$ which
contain infinite number of different lattice-points on $\SL_\RE$. 
In fact the power of $\MON$ can be not countable but continuous in general. 
We can write the elements of $\MON$ as 
$
l_n(r)=(N_r+n)\SE,
$ 
where $n$ can be elements of $\SN-\NA$, which 
satisfy the relation $n \SE\in {\rm Mon}(0)$. 
There are a lot of different possibilities depending on the choice of 
the original non-standard natural number $ ^*N \in\SN-\NA$. 
We shall here show two examples, that is, 
one has an infinite series of $\SM$ and the other a finite series.

\hfil\break
{\bf (1) Infinite series of $\SM$}

Define an infinite series of infinite non-standard natural numbers
by the following ultra-products;[1] 
\be
^*N_M\equiv \prod_{n\in\NA}\A_n^{(M)},\ \ \ {\rm for}\ \ M\in \NA 
\ee 
where $ \A_n^{(M)}=1$ for $0\leq n \leq M$ and $\A_n^{(M)}=(n+1)^{n-M}$ 
for $n>M$. 
Following the definition of the order $>$ for ultra-products, 
  we see that all of $^*N_M$ are infinity and the order 
is given by
$
^*N_0>\ ^*N_1>\ ^*N_2>\cdots.
$ 
Then we have an infinite series of infinitesimal numbers
$
\SE_0< \ \SE_1<\ \SE_2<\ \cdots,
$ 
where $\SE_M\equiv (^*N_M)^{-1}$. 
We can also prove that ratios 
\be
^*\lm_M\equiv {\ ^*N_{M-1} \over \ ^*N_M}, 
\ \ \ {\rm for}\ \ M\geq 1
\ee 
are infinities of $\SN$. 
Since $^*N_0$ is an element of natural numbers 
$\SN-\NA$, we can take 
$
\SE=\SE_0.
$ 
Here let us consider the following rescaling for the lattice points;
\be
l_n(r)-l_0(r)=n\SE_0\equiv \ ^*\lm_1^{-1}l_n^{(1)}, 
\ee 
where 
$
l_n^{(1)}=n\SE_1.
$ 
Note that $l_n^{(1)}$ is independent of $r$. 
Even if the relation $n\SE_0\in {\rm Mon}(0)$ must be satisfied, 
the set of $n \in \SN$ satisfying the relation 
contains non-standard integers 
\be
n_m^{(1)}\equiv m\times \ ^*N_1 \in \ ^*{\cal Z}, 
\ \ \ {\rm for}\ \ \forall m\in {\cal Z}. 
\ee 
It is trivial that the relation is satisfied as 
$
n_m^{(1)}\SE_0=m\lm_1^{-1}\in {\rm Mon}(0).
$ 
It is also obvious that 
$
n_m^{(1)}\SE_1=m\ \in {\cal Z}.
$ 
Thus we can see that the set of $\forall l_n^{(1)},\ \ \SL_\RE^{(1)}
\equiv \{l_n^{(1)}=n\SE_1|n\in \ ^*{\cal Z},\ n\SE_0\in {\rm Mon}(0)\}$, 
is an infinitesimal-lattice space with the lattice-length $\SE_1$. 
In fact the set $\SL_\RE^{(1)}$ is constructed from
the elements of $\MON$ rescaled by the factor $\ ^*\lm_1$.  
From the facts that $\SL_\RE^{(1)}$ contains all integers, Archimedian 
property certifies the existence of natural numbers $m \geq |r|$ for $\forall r
\in \RE$ and the distance between two neighboring lattice-points is 
an infinitesimal number $\SE_1$, we can find 
an element of $\SN-\NA$, $N_r^{(1)}$, satisfying the relation 
\be
N_r^{(1)}\SE_1\leq r^{(1)}<(N_r^{(1)}+1)\SE_1,\ \ \ {\rm for}\ \ 
\forall r^{(1)}\in\RE.
\ee 
Following the same argument for the construction of $\SM$, 
we can introduce the monad of $r^{(1)}$, Mon($r^{(1)}|\SL_\RE^{(1)}$), 
by the set of the following lattice-points on $\SL_\RE^{(1)}$;
\be
l_n^{(1)}(r^{(1)})=(N_r^{(1)}+n^{(1)})\SE_1,
\ee 
where $n^{(1)}\in \ ^*{\cal Z}$ and st$(n^{(1)}\SE_1)=0$ must be fulfilled. 
It is obvious that Mon($r^{(1)}|\SL_\RE^{(1)}$) contains an infinite number of 
elements. 
Now we can define $\SM^{(1)}$ by the set 
\be
\SM^{(1)}\equiv \SL_\RE^{(1)}/{\rm Mon}(0|\SL_\RE^{(1)}). 
\ee 
The relation 
\be
\SM^{(1)}\cong \SM\cong \RE
\ee 
as additive groups is obvious. 
Thus translations and rotations on $N$-dimensional space $(\SM^{(1)})^N$ 
are described as same as those of $(\SM)^N$. 
We can conclude that every monad lattice-space 
$\MON$ for $\forall r\in \RE$ contain the 
same space $\SM^{(1)}$ by means of the same scale transformation. 

By using the infinite series of $^*N_M$ we can proceed the same argument for the
construction of $\SM^{(M)}$ and thus we obtain the infinite series of 
sets isomorphic to $\RE$ as additive group such that 
$
\RE\cong \SM \cong \SM^{(1)} \cong \cdots \cong \SM^{(M)}\cong \cdots.
$ 

\hfil\break
{\bf (2) Finite series of $\SM$}

We definite a finite series of infinite numbers
\be
^*N_l^L\equiv \prod_{n\in \NA} (n+1)^{L-l}, \ \ \ {\rm for}\ l=0,1,2,\cdots,L-1
\ee 
where $^*N_l^L\in \SN-\NA$. 
We also see that 
\be
^*\lm_l\equiv {\ ^*N_{l-1}^L \over \ ^*N_l^L}=\prod_{n\in \NA}(n+1) \in \SN-\NA.
\ee 
Following the same argument as that of the infinite series, we can construct 
a finite series of sets isomorphic to $\RE$ as additive group
$ 
\RE\cong \SM\cong \SM^{(1)}\cong \cdots \cong \SM^{(L-1)}. 
$ 

\section{ Construction of fields on $\SM$}
Here we shall construct fields on $ ^*{\cal M}$. 
In the construction of field theory on $ ^*{\cal M}$ 
we follow the next two fundamental principles:[12,13,14]
\hfil\break
(I) All definitions and evaluations should be carried out 
on the original space $ ^*{\cal L}$. 
\hfil\break
(II) In definitions of any kinds of physical quantities on $\ ^*{\cal M}$, 
all the fields contained in the same monad lattice-space Mon$(r| ^*{\cal L})$ 
should be treated equivalently. (Principle of physical equivalece) 
\hfil\break 
It should be noted that the principle (I) means that theories which we will 
make on $\SL$ is generally not the same as any extensions of standard 
theories 
which have been constructed on $\RE$. 
The principle (I) also tells us 
that all physical expectation values on $\RE$ are obtained 
by taking standard part maps (maps from $ ^*{\cal R}$ to ${\cal R}$)[1] of 
results calculated on $ ^*{\cal L}$. 
The principle (II) is considered as the realization of 
the equivalence for 
indistinguishable quantities in quantum mechanics on non-standard space.[3] 
This principle, principle of physical equivalence, determines projections 
of physical systems defined on $\SL$ to those defined on $\SM$. 
Taking account of the fact that all points contained in 
the same monad lattice-space Mon$(r| ^*{\cal L})$ 
cannot be experimentally distinguished, the equivalent treatment with 
respect to all quantities defined on these indistinguishable 
 points is a natural requirement in the construction 
of theories on $ ^*{\cal M}$. 

\hfil\break
{\bf (1) Fields on $\SL$}

Let us define two fields $A(m)$ and $\bar A(m)$ on every lattice point 
$r(m)$ on $\ ^*{\cal L}$, which follow the commutation relations 
$
[A(m),\bar A(m')]=\delta_{mm'}\ \ \ {\rm and\ \  others}=0.
$ 
The vacuum $| ^*0>=\prod_m|0>_m$ and the dual vacuum ${ <\ ^*\bar 0|}
=\prod_m\ _m<\bar 0|$ 
are defined by 
$
A(m)|0>_m=0\ \ \ \ {\rm and} \ \ \ \  _m<\bar 0|\bar A(m)=0
$ 
with $_m<\bar 0|0>_m=1$. 
The fields $A(m)$ and $\bar A(m)$ operate only on the vacuum $|0>_m$ and the 
dual vacuum $_m<\bar 0|$. 
Following the principle (I), 
all expectation values are imposed to be calculated 
on $ ^*{\cal L}$ such that 
$
<\ ^*\bar 0|\hat {\cal O}(\{ A\}, \{\bar A\})| ^*0>\ \in \SR, 
$
where $\hat {\cal O}$ is operator constructed from the sets of the fields 
$A(m)$ and $\bar A(m)$. 
Physical values are obtained by the standard part map[1] as
$
{\rm st}(<\ ^*\bar 0| \hat {\cal O} | ^*0>) \in    {\cal R}.
$

\hfil\break
{\bf (2) Fields on $\SM$}

Following principle of physical equivalence (principle (II)), 
we define fields at every point on $ ^*{\cal M}$ 
as the following equivalent sum over all fields contained 
in Mon$(r| ^*{\cal L})$;
\be
\varphi([r])\equiv \ ^*\sum_{l} A(N_r+l)/\sqrt{^*\sum_{l}1}, \ \ \ \ \ 
\bar  \varphi ([r])\equiv \ ^*\sum_{l} \bar A(N_r+l)/
\sqrt{^*\sum_{l}1},
\ee 
where $^*\sum_l\equiv \sum_{l, ^*\varepsilon l\in {\rm Mon}(0)} $ and 
hereafter 
$[r]$ in $\varphi([r])$ always means the fact that the equivalent sum over 
Mon$(r| ^*{\cal L})$ 
expressed by $^*\sum_l$ is carried out in the definition of $\varphi ([r])$. 
Here the equivalent sum is just the expression of principle of physical 
equivalence. 
We can easily evaluate the commutation relation 
\be
[\varphi ([r]),\bar \varphi ([r'])]=\ ^*\delta_{rr'}=1 \ ({\rm for}\ \ r'=r),
  \ \ \              =0  \ ({\rm for}\ \ r'\not=r).
\ee 
Note that $r,r'\in {\cal R}$ but $ ^*\delta_{rr'}$ is not equal to the usual 
Dirac delta function $\delta(r-r')$. 
Complex fields on $ ^*{\cal M}$, which are represented by linear 
combinations certifying the same weight 
for all fields contained in Mon$(r| ^*{\cal L})$, are generally written by 
\bea 
\varphi ([r];k)&=&\ ^*\sum_{l} e^{i\theta_l^k(r)} A(N_r+l)/\sqrt{^*\sum_{l}1},\ \ \ \ \nonumber\\
\bar \varphi ([r];k)&=&\ ^*\sum_{ l}
e^{-i\theta_l^k(r)} \bar A(N_r+l)/\sqrt{^*\sum_{l}1},
\eea 
where
$
\theta_l^k(r)=\theta_k(r)+2\pi l k/\ ^*\sum_{l}1
$ 
with the constraint  $ ^*\varepsilon k\in {\rm Mon}(0)$ for 
non-standard integers $k$. 
They satisfy the commutation relations 
\be 
[\varphi ([r];k),\bar \varphi ([r'];k')]=\ ^*\delta_{rr'}\delta_{kk'}
\ \ 
{\rm and \ \ others}=0. 
\ee 
These fields are the Fourier components for the fields on Mon$(r| ^*{\cal L})$ 
and their component number is same as that of $A(N_r+l)$ and $\bar A(N_r+l)$ 
included in Mon$(r| ^*{\cal L})$, because the constraint for $k$, that is, 
$ ^*\varepsilon k\in {\rm Mon}(0)$, is same as that for $l$. 
Note that $\varphi ([r])$ and $\bar \varphi ([r])$ 
correspond to the above fields with $k=0$ and $\theta_0(r)=0$. 
Experimentally the differences of the wave numbers $k$ are not observable, 
because their wave lengths are infinitesimal. 
It is stressed that fields on one point of $\SM$ have infinite degrees of 
freedom. 
General fields on $\SM$ are described by functions of these fields 
$
\phi([r])=f(\{\varphi ([r];k)\}, \{\bar \varphi ([r];k)\}). 
$ 

\hfil\break
{\bf  (3) Extension to $N$-dimensional space} 

Extension of the fields to $N$-dimensional space $(^*{\cal M})^N$ 
is trivial. 
Note that one should not confuse the $N$ for the $N$-dimensions of the space 
with the $N_{r}$ for the lattice number corresponding to $r$ of $\RE$ 
in the discussions. 
Every point of $( ^*{\cal L})^N$ is represented by a $N$-dimensional 
vector 
$
\vec r^N(\vec m)\equiv(r_1(m_1),.....,r_N(m_N)), 
$ 
where $r_i(m_i)=\ ^*\varepsilon(N_{r_i}+l_i)$ with st$( ^*\varepsilon N_{r_i})=r_i \in 
{\cal R}$ and $ ^*\varepsilon l_i \in {\rm Mon}(0)$ for $i=1,.....,N$. 
Fields with $N$-components at $\vec r^N(\vec m)$, 
$A_j(\vec m)$ and $\bar A_k(\vec m)$ $(j,k=1,....,N)$, 
are defined by the commutation relations 
\begin{equation}
[A_j(\vec m), \bar A_k(\vec m')]=\delta_{jk}\prod_{i=1}^N\delta_{m_im'_i} 
\ \ ({\rm for}\ \ j,k=1,....,N)\ \ {\rm and \ \ others}=0.
\end{equation}
We may consider that these $N$ number of fields describe $N$ 
oscillators of a lattice point corresponding to $N$ dimensional space-time 
axes. 
The fields on $( ^*{\cal M})^N$ are described as 
\begin{equation}
\varphi_j ([\vec r^N];\vec k^N)=\ ^*\sum_{l_1}\cdot\cdot\cdot
\ ^*\sum_{l_N} e^{i\sum_{s=1}^N\theta_{l(s)}^{k(s)}
(\vec r^N)} A_j(N_{r_1}+l_1,\cdot\cdot\cdot,N_{r_N}+l_N)/( ^*\sum_l 1)^{N/2}
\end{equation} 
and similar to $\bar \varphi _j([\vec r^N];\vec k)$. 
We again have the commutation relations
\be
[\varphi _j([\vec r^N];\vec k^N), \bar \varphi _l([\vec r'^N];\vec k'^N)]=
\delta_{jl}\prod_{i=1}^N( ^*\delta_{r_ir'_i}\ \delta_{k_ik'_i}) \ \ 
{\rm and\ \  others}=0. 
\ee


\section{ Internal symmetries on $(\SM)^N$} 
Symmetries  
induced from the internal substructure 
$({\rm Mon}(r| ^*{\cal L}))^N$ on $( ^*{\cal M})^N$ are expressed by 
transformations $U_T$ which keep all expectation values unchanged such that
$$
<\ ^*\bar 0|\hat {\cal O}(\{A\},\{\bar A\})| ^*0>
=<\ ^*\bar 0|U_T^{-1} U_T\hat {\cal O}(\{A\},\{\bar A\})U_T^{-1} U_T| ^*0>. 
$$ 
In general the transformation $U_T$ will be represented by maps 
of fields $A_j(\vec m)\ (\bar A_j(\vec m))$ 
to a linear combination of the fields $A_k(\vec m)\ (\bar A_k(\vec m))
\ (k=1,\cdots,N)$ on $\SL$. 
If the operators $U_T$ do not change the structure of $(\SM)^N$, 
they can represent symmetries on $(\MON)^N$.  

\hfil\break
{\bf  (1) Transformation opertors on internal subspaces $(\MON)^N$}

Let us start from the construction of 
transformation operators on an internal subspace contained in 
a point on $( ^*{\cal M})^N$ corresponding to a point 
$\vec r^N=(r_1,.....,r_N)$ on $\RE^N$. 
The transformations map 
fields $A_j(\vec r^N(\vec m))$ 
($\bar A_j(\vec r^N(\vec m))$) 
on every lattice-point 
($\vec r^N(\vec m)=(N_{r_1}+l_1,...,N_{r_N}+l_N)$) 
to linear combinations of fields 
$A_k(N_{r_1}+l_1',...,N_{r_N}+l_N')$ 
($\bar A_k(N_{r_1}+l_1',...,N_{r_N}+l_N')$) ($k=1,...,N$) 
on the lattice-points of the same subspace. 
Following principle of physical equivalence (principle (II)), we construct 
the following $N^2$-number of operators $\hat T_{jk}([\vec r^N])$ 
on $(\SM)^N$, which are again defined by the 
equivalent sum over all fields contained in the $N$-dimensional subspace 
$({\rm Mon}(r| ^*{\cal L}))^N$ as 
\begin{equation}
\hat T_{jk}([\vec r^N])
=\ ^*\sum_{l_1}\cdot\cdot\cdot\ ^*\sum_{l_N} 
\bar A_j(N_{r_1}+l_1,...,N_{r_N}+l_N)
                                A_k(N_{r_1}+l_1,...,N_{r_N}+l_N).
\end{equation} 
We easily obtain commutation relations 
$$ 
[\hat T_{jk}([\vec r^N]),A_l(\vec r'^N(\vec m))]=
-(\prod_{i=1}^N\ ^* \delta_{r_ir'_i})\delta_{jl}A_k(\vec r^N(\vec m)), 
$$ 
$$ 
[\hat T_{jk}([\vec r^N]),\bar A_l(\vec r'^N(\vec m))]=
(\prod_{i=1}^N\ ^* \delta_{r_ir'_i})\delta_{kl}\bar A_j(\vec r^N(\vec m)), 
$$ 
\be
[\hat T_{jk}([\vec r^N]),\hat T_{lm}([\vec r'^N])]
=(\prod_{i=1}^N\ ^* \delta_{r_ir'_i}) (\delta_{kl}\hat T_{jm}([\vec r^N])
                                  -\delta_{jm}\hat T_{lk}([\vec r^N])).
\ee 
These operators $\hat T_{jk}$ can be recomposed into the following generators; 
\hfil\break
(1) $U(1)$-generator:$\ \ \ 
\hat J_0=\sum_{j=1}^N \hat T_{jj}.
$ 
\hfil\break 
(2) $SU(N)$-generators:$\ \ \ 
\hat J_L=\sum_{j=1}^{L+1}g_j \hat T_{jj},\ \  \ \ {\rm for}\ \ L=1,...,N-1
$ 
with the traceless condition $\sum_{j=1}^{L+1}g_j=0$, and 
$ 
\hat J_{jk}^{(1)}=\hat T_{jk}+ \hat T_{kj}\ \ {\rm and} \ \  
\hat J_{jk}^{(2)}={1 \over i}(\hat T_{jk}- \hat T_{kj}) 
\ \ {\rm for}\ \ j\not=k. 
$ 
Now it is trivial that operators given by 
\begin{equation}
U(\{\alpha(\vec r^N)\})={\rm exp}[i\sum_{j=1}^N\sum_{k=1}^N \alpha_{jk}(\vec r^N)\hat T_{jk}
([\vec r^N])]
\end{equation}
with st(
$
\forall \alpha_{jk}(\vec r^N))\in\ {\cal C}
$ 
(the set of complex numbers) produce maps of all fields on the subspace 
$(\MON)^N$ to linear combinations of the fields 
on the same subspace. 
From the construction procedure of $\hat T_{jk}$ 
it is obvious that the operators do not break the structure of $(\SM)^N$. 
Note also that $U$ does not change the vacuum and the dual vacuum, because 
$
\forall \hat T_{jk}| ^*0>=< ^*\bar 0|\forall \hat T_{jk}=0.
$

\hfil\break
{\bf (2) Symmetries on $(\SM)^N$}

Operators on $(\SM)^N$ can be defined by products of $U(\{\A(\vec r^N)\})$ 
as
\be
U_T(\{\A\})=\prod_{i=1}^N \ ^*\prod_{N_{ri}}U(\{\A(\vec r^N)\}), 
\ee 
where $\ ^*\prod_{N_{ri}}$ stand for the product with respect to 
$\forall N_{r_i}$ 
with the constraint st$(\SE N_{r_i})=r_i\in\RE$ . 
It is interesting that the transformations produced by $U_T(\{\alpha\})$ 
are generally local transformations on our observed space $( ^*{\cal M})^N$ 
because the parameters $\{\alpha\}$ can depend on the position $\vec r^N$,   
whereas they are global ones on the internal subspace 
$({\rm Mon}(r| ^*{\cal L}))^N$. 
Note that $U_T$ does not change the vacuum and the dual vacuum.

Let us show a few realistic transformations included in $U_T$. 
\hfil\break
(a) $U(1)$ transformation:
\begin{equation}
U_0(\vec r^N)={\rm exp}[i\alpha_0(\vec r^N)\hat J_0([\vec r^N])]
\end{equation} 
${\rm for\ \  st}(\alpha_0)\in  {\cal R}$. 
It is an interesting problem to investigate whether  this $U(1)$ symmetry can be 
the $U(1)$ symmetry of electro-weak gauge theory or the solution of so-called 
$U(1)$ problem in hadron dynamics. 
\hfil\break 
(b) $SU(N)$ transformation:
\begin{equation}
U_N(\vec r^N)={\rm exp}[i\{\sum_{L=1}^{N-1}\alpha_L(\vec r^N)\hat J_L([\vec r^N])
                +\sum_{j=1}^{k-1}\sum_{k=2}^{N}\sum_{i=1}^2\alpha_{jk}^{(i)}(\vec r^N)
                  \hat J_{jk}^{(i)}([\vec r^N])\}]
\end{equation} 
for st$(\forall \alpha_L), \ {\rm st }( \forall \alpha_{jk}^{(i)}) \in {\cal R}$. 
It is an interesting proposal 
that three color components of QCD may be identified by those of $U_3(\vec r^3)$
for three spatial dimensions.   


\section{ Quantized configuration space} 
In usual field theory space-time variables are treated as parameters.  
Here we construct configuration space describing 
$ ^*{\cal M}$, where the space-time are expressed by operators. 

\hfil\break
{ \bf (1) Quantization of configuration space}

We can construct  position operator for 1-dimensional space
\be 
\hat r_{\SM}=\ ^*\sum_{N_r} r \hat T_r,
\ee 
where $\ ^*\sum_{N_r}$ stands for the sum over $\forall N_r$ with the 
constraint st$(\SE N_r)=r\in\RE$ and 
$
\hat T_r=\ ^*\sum_l \bar A(N_r+l)A(N_r+l).
$ 
Following 
principle of physical equivqlence, 
$\hat T_r$ is expressed by the equivalent sum with respect to 
all fields in the same monad lattice-space $\MON$.  
Note that $r$ in (77) can be replaced by $r+a_r \SE$ with the constant 
st$(a_r \SE)=0$ for $\forall r\in \RE$. 
The eigenstate of $\hat r_{\SM}$ for the eigenvalue $r$ is written by 
\be 
|r>_{\SM}\equiv  \bar \varphi([r])
| ^*0>. 
\ee 
Hereafter we call them monad states. 
The relation
$
\hat r_{\SM}|r>_{\SM}=r|r>_{\SM}
$ 
is trivial. 
If one does not want to have 0 eigenvalue for $r=0$, $r+a_r \SE$ can be used 
instead of $r$ in the definitoin of $\hat r_{\SM}$. 
The monad states $|r>_{\SM}$ are quite similar to the ket states 
of usual quantum mechanics except the normalization condition 
$
\ _{\SM}<r|r'>_{\SM}=\ ^*\delta_{rr'}, 
$ 
where $_{\SM}<r|=<\ ^*\bar 0|\ ^*\prod_{N_r}\varphi([r])$. 
It is noted that every monad state $|r>_{\SM}$ has its own 
internal substructure $\MON$. 

Now we can define the quantized states for our configuration space as follows;
\be 
|\SM>\equiv \ ^*\prod_{N_r} |r>_{\SM},\ \ \ <\SM|\equiv \ ^*\prod_{N_r}\ _{\SM}<r|.
\ee 
On these states the position operator $\hat r_{\SM}$ is represented 
by a diagonal operator 
and then we can consider that the base state $|\SM>$ describes our configuration space, which is normalized as 
$<\SM|\SM>=1$. 

Extension to $N$-dimension is trivial. 
A component of the position-vector operator can be defined as same as 
that of the 1-dimensinal case, e.g., for the $i$th component 
\be 
\hat r_{i\SM}=\ ^*\sum_{N_{r1}}\cdots\ ^*\sum_{N_{rN}} r_i 
\hat T_i([\vec r^N]),
\ee 
where 
$ 
\hat T_i([\vec r^N]) =\ ^*\sum_{l_1}\cdots\ ^*\sum_{l_N} 
\bar A_i(N_{r_1}+l_1,...,N_{r_N}+l_N)A_i(N_{r_1}+l_1,...,N_{r_N}+l_N) 
$ 
for $i=1,2,...,N$. 
The $N$-dimensional configuration state is expressed by 
$ 
|\SM^N>=\prod_{j=1}^N(\ ^*\prod_{N_{r1}}\cdots\ ^*\prod_{N_{rN}}
 \bar \varphi_j([\vec r^N]))| ^*0>. 
$

\hfil\break
{\bf (2) Infinitesimal distance}

We can define infinitesimal relative distance operators only on the internal 
subspace Mon$(r| ^*{\cal L})$ such that
\begin{equation}
d\hat r(\Delta l)\equiv \hat r(N_r+l)-\hat r(N_r+l'),
\end{equation} 
where 
$ \Delta l\equiv l- l'$ and 
$ 
\hat r(N_r+k)\equiv\ ^*\varepsilon (N_r+l)  \bar A([r]) A(N_r+k)
$ 
with the definiion 
$ 
\bar A([r])\equiv \ ^*\sum_l \bar A(N_r+l),
$ 
which follows princilpe of physical equivalnce. 
The monad states 
$|r>_{\SM}$ are the eigenstates of $\hat r(N_r+l)$ and $d\hat r(\Delta l)$. 
We actually obtain 
\begin{equation}
d\hat r(\Delta l)|r>_{\SM}=\ ^*\varepsilon \Delta l|r>_{\SM}. 
\end{equation}

We can write squared distance operators in the $N$-dimensional space  as 
$
(d \hat s)^2(\vec r^N)=d \hat r_\mu (\Delta \vec l^N) g^{\mu\nu}d \hat r_\nu
(\Delta \vec l^N),
$ 
where the sums over $\mu$ and $\nu$ from $1$ to $N$ are neglected, 
$$
d \hat r_\mu (\Delta \vec l^N)=
\hat r_\mu(N_{r_1}+l_1,...,N_{r_N}+l_N)-
\hat r_\mu(N_{r_1}+l'_1,...,N_{r_N}+l'_N)
$$ with 
$\hat r_\mu(N_{r_1}+l_1,...,N_{r_N}+l_N)=\ ^*\varepsilon (N_{r_\mu}+l_\mu)
\bar A_\mu([\vec r^N])A_\mu(N_{r_1}+l_1,...,N_{r_N}+l_N)$ 
and 
$
\Delta \vec l^N=(l_1-l'_1,\cdot\cdot\cdot,l_N-l'_N). 
$
If the metric oprator $g^{\mu\nu}$ is taken as Minkowski metric,
the internal subspace $({\rm Mon}(r| ^*{\cal L}))^N$ 
just represents so-called local inertial system in general relativity. 
We have the equations 
\bea 
d \hat r_\mu (\Delta \vec l^N)|\vec r^N>_{\SM}&=&
\SE\Delta l_\mu |\vec r^N>_{\SM}, \CR
(d\vec s)^2(\vec r^N)|\vec r^N>&=&\SE^2 \Delta l_\mu g^{\mu\nu} \Delta l_\nu
|\vec r^N>_{\SM}. 
\eea 
The expectation value of $(d \hat s)^2$ is calculated as 
$
(ds)^2=_{\SM}<\vec r^N|(d \hat s)^2(\vec r^N)|\vec r^N>_{\SM}.
$ 
The same expectaton value of the squared distance operator can be 
obtained in terms of the expectation value 
with respect to the configuration state 
$|\SM^N>$. 
It is transparent 
that transformations keeping $(ds)^2$ unchanged 
are represented by $U(\{\alpha(\vec r^N)\})$.

\section{ Translations, Rotations and 
relativistic transformations}
We shall study symmetries on the configuration space, 
which keep all expectation values  
unchanged such that 
$
<\SM^N|U^{-1}U \hat {\cal O}(\{\bar A\},\{A\})U^{-1}U|\SM^N>.
$ 
Note that the configuration state $|\SM^N>$, the dual state $<\SM^N|$ 
and operators 
are transformed as 
$$
|\SM^N>\longrightarrow U|\SM^N>,\ \ \ 
<\SM^N|\longrightarrow <\SM^N|U^{-1},\ \ \  
U \hat {\cal O}(....)U^{-1}.
$$ 

\hfil\break
{\bf (1) Translational invariance on $(\SM)^N$}

The operator which replaces $|r>$ with $|r+\Delta>$ for $\Delta \in \RE$ 
is obtained as 
\be
\hat p_r(\Delta)=\ ^*\sum_l \bar A(N_{r+\Delta}+l)A(N_r+l). 
\ee 
We have $\hat p_r(\Delta)| ^*0>=0$. 
Then we can define the translation operator  by 
\be
\ \ \ \ \hat P(\Delta)= :\ ^*\prod_{N_r} \hat p_r(\Delta):,\ \ \ \ 
\ee 
where $:......:$ means the normal product used in usual field theory, 
in which all creation operators ($\bar A_j(m)$) must put on the left-hand 
side of all annihilation operators ($A_j(m)$). 
We see that $\hat P(\Delta)$ transforms the configuration state $|\SM>$ 
to the isomorophic space 
$ 
\hat P(\Delta)|\SM>\cong |\SM>
$ 
for st$(\forall \Delta) \in \RE$. 

Let us study the invariance of expectation values 
$
<\SM|\hat {\cal O}(\{\bar A\}, \{A\})|\SM>. 
$ 
Taking account of the definitions of
$|\SM>=\prod \bar \varphi ([r])|\ ^*0>$ and 
$<\SM|=<\ ^*\bar 0|\prod \varphi([r])$ 
and the fact that 
all the fields commute each other except $A$ and $\bar A$ on the same 
lattice-point, the number of $A$ and that of $\bar A$ 
on the same lattice-point must be same in operators having non-vanishing 
expectation values on $|\SM>$. 
This means that  every term of such operators 
must be written by 
the product of powers 
such as $(\bar A A)^n$ with $n\in \NA$ for all pairs of $A$ and 
$\bar A$ on the same lattice-point. 
On the other hand we easily see that the products of $\bar A A$ on the same 
lattice-piont commute with $\hat P(\Delta)$ such that 
$ 
[\bar A A,\hat P(\Delta)]=0
$ 
for $\forall \Delta \in \RE$. 
Now we can conclude that operators having non-vanishing expetation values 
commute with the translation operators, that is, 
$ 
[\hat {\cal O}(\{\bar A\}, \{A\}),\hat P(\Delta)]=0. 
$ 
Translational invariance is certified for physically meaningful operators 
as 
\be 
<\SM|\hat P(-\Delta)\hat {\cal O}(...)\hat P(\Delta)|\SM>
=<\SM|\hat {\cal O}(...)|\SM>, 
\ee 
where the commutatibity of $\hat {\cal O}$ and $\hat P$ and  
$<\SM|\hat P(-\Delta) \hat P(\Delta)=<\SM|$ 
are used. 

The extension of the above argument to the $N$-dimensional spaces is 
trivial. 

\hfil\break
{\bf (2) Rotations}

Rotational invariance can be introduced only for subspaces whose metric 
$g^{\mu\nu}$ have the same sign like $SO(3)$ subspace of 
$SO(3,1)$. 
Generators for the rotations in $(j,k)$-plane are given by 
$ 
\hat J_{jk}=\hat T_{jk}-\hat T_{kj}. 
$ 
In general rotation operators are described by 
\be 
U_R(\{\theta\})=e^{i\sum_{(j,k)}\theta_{jk} \hat J_{jk}}. 
\ee 
We see that $U_R$ for st$(\forall \theta_{jk}) \in \RE$ 
are unitary and generate rotations 
on the subspace. 

\hfil\break
{\bf (3) Lorentz transfomations}

Position operator for one point on $(\SM)^N$ corresponding to 
$\vec r^N$ on $\RE^N$ is given by 
\be 
\hat r_j([\vec r^N])=r_j \bar \varphi_j([\vec r^N]) \varphi_j([\vec r^N]), 
\ \ {\rm for}\ j=1,...,N. 
\ee 
The expectation value of squared distance from the origin are 
evaluated as 
$ 
(\vec r^N)^2=<\SM^N|\hat r_\mu ([\vec r^N])g^{\mu\nu}\hat r_\nu([\vec r^N]) 
|\SM^N>, 
$ 
where the metric tensors $g^{\mu\nu}$ are taken as Minkowski metric tensors. 
Let us study the simplest case for $N=2$. 
The metric tensors are chosen such that 
$
g^{11}=-g^{22}=1 \ \ {\rm and}\ \ g^{12}=g^{21}=0.
$ 
Transformations 
\be
U_L(a)=\prod_{j=1}^N \ ^*\prod_{N_{rj}} e^{-a\hat J_{12}^{(1)}([\vec r^N])}
\ee 
with the constraint st$(a) \in \RE$ 
generate 2-dimensional Lorentz 
transformations which are expressed in 2-dimensional matrices as 
\[   U_L(a)=
\left(
   \begin{array}{cc}
          \ \ {\rm cosh}a & \ \ -{\rm sinh}a   \\
           \ -{\rm sinh}a & \ \ \ \ {\rm cosh}a 
   \end{array}
\right)
\]

Generalization for the $N$-dimensions can be performed by using 
combinations of $U_L(a)$ with the rotations.

\hfil\break
{\bf (4) General relativistic transformations} 

We have many different types of transformations which 
keep the squred distance $(\vec r^N)^2$ invariant but generally do not 
the metric tensors invariant, while Lorentz transformations keep both 
of them invariant. 
They are described by the transformations $U_T(\{\A\})$, where 
the parameters $\{\A\}$ should be chosen such that all the axes are real 
after the translations.  
Of course, all the parameters must be finite. 
In such transformations we have different types of vectors corresponding 
to covariant and contravariant tensors in general coordinate transformations. 
The difference between them is expressed 
as follows; 
\bea
U_G \hat r_\mu|\SM^N>, &\ \ &{\rm for\  covariant\ vectors}  \CR
<\SM^N|\hat r_\mu g^{\mu\nu}U_G^{-1}, &\ \ &{\rm for\ contravariant\ vectors}.
\eea 

A simple example representing dilatation transformations 
are described by 
\be 
D_d=e^{\sum_{j=1}^N a_j(\vec r^N)\hat T_{jj}([\vec rN])} 
\ee 
with $\forall a_j(\vec r^N) \in \SR$, which transforms as
$$
U_d\ \hat r_\mu|\SM^N>=e^{a_{\mu}(\vec r^N)}\hat r_\mu|\SM^N>,
$$ 
$$
<\SM^N|\hat r_\nu g^{\nu\mu}U_d^{-1}=<\SM^N|\hat r_\nu g^{\nu\mu}
e^{-a_\mu(\vec r^N)}. 
$$ 
We see that these transformations change the eigenvalues of the covariant and 
the contravariant vectors.

Note that $U_G(\{\alpha(\vec r^N)\})$ is global on the subspace 
$(\MON)^N$, 
while it is generally local on observed space $(\SM)^N$. 
Note also that all the transformations described by $U_T$ 
can include general relativistic transformations. 
This fact implies that  general relativstivc transformations 
are generally represented by local non-abelian transformations.


\section{Concluding remarks }
We shall briefly comment 
that, instead of bosonic fields $A(m)$ and $\bar A(m)$, 
we can construct similar field theory 
by using fermionic fields $C(m)$ and $\bar C(m)$ which satisfy anticommutaion 
relations $[C(m),\bar C(m)]_+=1$ and commutation relations 
$[C(m),C(m')]_-=[C(m),\bar C(m')]_-
=[\bar C(m),\bar C(m')]_-=0$ for $m\not=m'$. 
As far as operators 
$\hat T_{jk}([\vec r^N])$ are 
concerned, we can define them by the replacement of $A$ and $\bar A$ with 
$C$ and $\bar C$, respectively. 
And we get the same commutation relations. 
This means that all the arguments of the internal 
symmetries performed in the bosonic oscillator case 
are completely accomplished in the fermionic oscillator case. 
That is to say, as far as the internal symmetries are concerned, 
there is no difference between the bosonic and the fermionic cases. 
Futhermore we can easily understand that not only $U_T$ but also all 
other operators written by the products 
of $\bar A$ and $A$ like $ \hat T_r,\ \hat r$ and $\hat p_r$ can be defined 
in the replacement of $\bar A\ A$ with $\bar C\ C$ and they have the same 
properties as discussed in the bosonic case. 
Difference between them appears in the construction of realistic fields 
from $\varphi([r];k)$. 
Namely products of more than the non-standard natural number 
$\ ^*\sum_l 1$ with respect to the fields 
$\varphi([r];k)$ vanish for the fermionic case, 
whereas there is no such restriction in the bosonic case. 
We may say that the concept of antiparticles will be introduced more easily 
in the fermionic case by using occupaton and unoccupation numbers of 
lattice-points of the monad lattice-space $\MON$. 
Anyhow the selection of the bosonic or the fermionic 
or both like supersymmetric 
is still open question at present.

We have constructed a field thoery on the quantized space-time by using 
infinitesimal-lattice space $(\ ^*{\cal L})^N$. 
In this scheme the internal subspace $({\rm Mon}(r| ^*{\cal L}))^N$ and 
the symmety transformation $U_T$ 
induced from the subspace are uniquely determined, when we 
construct the field theory on $( ^*{\cal M})^N\cong {\cal R}^N$. 
Since all definitions and evaluations are imposed to be done on $(\ ^*{\cal L})^N$, 
we can perform them in terms of $*$-finte sum in non-standard analysis. 
In fact we need not introduce any Dirac $\delta$-functions. 
In this scheme we can carry out all evaluations on configuration space, 
not on Fock space in usual field theory. 
This fact is an interesting advantage in the investigation of quantum gravity, 
as was seen in the introduction of the infinitesimal relative distance and 
the local inertial system. 
In order to investigate this model in more detail an inevitable problem is 
introducing equation of motions on $( ^*{\cal M})^N$, 
which will be represented by difference equation on Mon$(r| ^*{\cal L})$. 
It is also interesting to study relations between the general field $\phi([r])$ 
and observed fields like leptons, quarks, gauge fields and etc. 

Finally I would like to present the global view of theory on non-standard 
space once more. 
The fundamental concept is introducing the equivalence based on 
experimental errors (physical equivalence) into theories 
in a mathematically consistent logic, which is allowed only on non-standard 
spaces. 
On the spaces the physical equivalence determine projections from 
non-standard spaces to observed spaces isomorphic to $\RE^N$, 
which are described by filters in non-standard theory. 
In fact the filters determine topologies, because they determine 
the structure of the monad space and then that of the observed space.[1] 
We have to understand that in an experiment
 we are allowed to peep only through 
a filter which is determined by the physical 
equivalence based on the errors of the experiment.  
Theories on observed spaces, which explain experimental results, of course 
have to depend on 
the filters which determine the projections of the theory on the 
non-standard space to 
theories on the observed spaces, 
even if the theory is uniqe on the non-standard space. 
Different filters derive different monad spaces and then 
different observed spaces (different theories). 
I would like again to repeat that we cannot perform any expriments 
which are not accompanied by any errors. 
Therefore we have always to take account of phenomena hidden behind 
experimental errors, when we make theories in our observed spaces.

\hfil\break
{\large {\bf References}}
\vskip1pt
\hfil\break
[1] A. Robinson, {\it Non Standard Analysis,} (North-Holland, Amsterdam, 1970).
\hfil\break
[2] T. Kobayashi, {\it Nuovo Cim.,} {\bf 113B } (1998) 1407.
\hfil\break
[3] T. Kobayashi, {\it Proceedings of 5th Wigner Symposium,} edited by 
P. Kasperkovitz and D. Grau (World Scientific, Singapoe, 1998) 518.
\hfil\break
[4] T. Kobayahi, {\it Symmetries 
in Science X,} edited by B.Gruber and M. Ramek (Plenum Press, New York and 
London, 1998) 153.
\hfil\break
[5] M. O. Farrukh, {\it J. Math. Phys.,} {\bf 16} (1975) 177.
\hfil\break
[6] T. Kobayashi, {\it Symmetries in Science VII,} edited by B.Gruber and T. 
Otsuka, (Plenum Press, New York, 1994) 287.
\hfil\break
[7] T. Kobayashi, {\it Nuovo Cim.,} {\bf 110B} (1995) 61.
\hfil\break
[8] T. Kobayashi,  {\it Nuovo Cim.,} {\bf 111B} (1996) 227.
\hfil\break
[9] T. Kobayashi, {\it Proceedings of the Fourth International 
Conference on Squeezed States and Uncertainty Relations,} edited by A. Han, 
(NASA Conference Publication 3322, 1996) 301. 
\hfil\break
[10] T. Kobayashi, {\it Phys. Lett.,} {\bf A 207} (1995) 320; {\bf A 210} (1996) 241; {\bf A 222} (1996) 26.
\hfil\break
[11] T. Kobayashi, {\it Translastions, Rotations and Confined Fractal 
Property on Infinitesimal-Lattice Spaces}, preprint of University of Tsukuba 
(1997). 
\hfil\break
[12] T. Kobatashi, {\it Talk in  the XI International Conference on Problems of Quantum 
Field Theory,} July, 1998, Dubna, Russia (to appear in the Proceedings). 
\hfil\break
[13] T. Kobayashi. {\it Talk in the International Symposium on Symmetries in 
Science XI,} August, 1999, Bregenz, Austria. 
\hfil\break
[14] T.Kobayashi, {\it Physical Equivalence on Non-Standard Spaces and 
Symmetries on Infinitesimal-Lattice Spaces,} preprint of Tsukuba-College of Technology, hep-th/9910152 (1999). 

\end{document}